\documentclass[11pt]{article}
\pdfoutput=1
\usepackage{eacl2014}
\usepackage{times}
\usepackage{url}
\usepackage{latexsym}
\usepackage{url}
\usepackage[fleqn]{amsmath}
\usepackage{graphicx}   
\usepackage{amsfonts}
\usepackage{color}
\title{Squeezing bottlenecks: exploring the limits of autoencoder semantic representation capabilities}

\author{Parth Gupta$^1$, Rafael E. Banchs$^2$ and Paolo Rosso$^1$ \\\\
  $^1$Natural Language Engineering Lab- PRHLT \\
Department of Information Systems and Computation \\
    Universitat Polit\`ecnica de Val\`encia, Spain \\
    {\tt \{pgupta,prosso\}@dsic.upv.es} \\\\
    $^2$Institute for Infocomm Research, Singapore\\
  {\tt rembanchs@i2r.a-star.edu.sg}  \\}

\begin{document}
\maketitle
\begin{abstract}
We present a comprehensive study on the use of autoencoders for modelling text data, in which (differently from previous studies) we focus our attention on the following issues: \textit{i)} we explore the suitability of two different models bDA and rsDA for constructing deep autoencoders for text data at the sentence level; \textit{ii)} we propose and evaluate two novel metrics for better assessing the text-reconstruction capabilities of autoencoders; and \textit{iii)} we propose an automatic method to find the critical bottleneck dimensionality for text language representations (below which structural information is lost).

\end{abstract}

\section{Introduction}
One of the major hurdles in comparing text in vector space models (VSM) is to deal with problems like \textit{synonymy} and \textit{polysymy}.
Usually in vector space, the documents are composed of thousands of dimensions.
In addition to high computational complexity, many meaningful associations between terms are shadowed by large dimensions.
There are models which try to solve this problem \textit{e.g.} pseudo relevance feedback (PRF) and explicit semantic analysis (ESA)~\cite{xu:1996,gabrilovich:2007}.
Other category of attempts to solve this problem comprise of dimensionality reduction techniques.

The goal of dimensionality reduction techniques is to transform high dimensional data ($\mathbb{R}^n$) into a much lower dimension representation ($\mathbb{R}^m$) pertaining the inherent structure of the original data where $m<<n$.
One such widely used approach is latent semantic indexing (LSI) which extracts a low rank approximation of a term-document matrix by means of principal component analysis (PCA)~\cite{deerwester:1990}.
There are some advanced approaches like probabilistic latent semantic analysis (PLSA) and latent dirichlet allocation (LDA) which observe the distribution of latent topics for the given documents~\cite{hofmann:1999,blei:2003}.

Dimensionality reduction techniques are also prominent while estimating the similarity between text across languages.
Associations of terms and documents across languages in such techniques are learnt by means of parallel or comparable text~\cite{nie:1999,banchs:2008,platt:2010}.

Dimensionality reduction techniques can broadly be categorised in two classes: linear and non-linear.
Usually, non-linear techniques can find more compact representations of the data compared to their linear counterparts~\cite{hinton:2006}.
If there exists statistical dependence among the principal components of PCA, or principal components have non-linear dependencies, PCA would require a larger dimensionality to properly represent the data when compared to non-linear techniques.

On the other hand, although non-linear projection methods such as multidimensional scaling (MDS) give a way to obtain much better representations for mono and cross-language similarity estimation, it is a transductive method~\cite{cox:2001,banchs:2008}.
It means MDS does not provide an operator to project the unseen data into the target low dimensional space like the resulting projection matrix in the case of PCA.

Lately, dimensionality reduction techniques based on deep-learning have become very popular, especially deep autoencoders (DA).
Deep autoencoders can extract highly useful and compact features from the structural information of the data.
Deep autoencoders have proven to be very effective in learning reduced space representations of the data for similarity estimation, \textit{i.e.} similar documents tend to have similar abstract representations~\cite{hinton:2006,salakhutdinov:2009b}.
Deep learning is inspired by biological studies which state the brain has a deep architecture.
Despite their high suitability to the task, deep-learning did not find much audience because of convergence issues until Hinton and Salakhutdinov~\shortcite{hinton:2006} gave a way to initialise the network parameters in a good region for finding optimal solutions.

Deep learning based dimensionality reduction techniques are quite popular for NLP tasks like sentiment prediction~\cite{socher:2011}, part-of-
speech tagging, chunking, named entity recognition, semantic role labeling~\cite{collobert:2011} and semantic composionality~\cite{socher:2012,socher:2013}. 
Unlike such tasks where sequential information among the words is important, 
information retrieval like similarity estimation based models such as ours use bag-of-words representation of the  text~\cite{yih:2011,platt:2010,salakhutdinov:2009b}.

Although deep learning techniques are in vogue, there still exist some important open questions.
In most of the studies involving the use of these techniques for dimensionality reduction, the qualitative analysis of projections is never presented.
This makes the assessment of the reliability of learning very difficult.
Typically, the reliability of the autoencoder is estimated based on its reconstruction capability. 

The objective of this work is to propose a novel framework for evaluating the quality of the dimensionality reduction task based on the merits of the application under consideration: the representation of text data in low dimensional spaces.

Concretely, our proposed framework is comprised of two metrics, 
\textit{structure preservation index (SPI)} and
\textit{similarity accumulation index (SAI)},
which capture different aspects of the autoencoder's reconstruction capability like the structural distortion of the data and similarities among the reconstructed vectors. In this way, our proposed framework gives better insight of the autoencoder performance allowing for conducting better error analysis and evaluation, and, as explained below, these metrics also provides a better means for estimating the adequate size of critical bottleneck dimensions.

We carry out the experiments of dimensionality reduction of text at sentence level and assess  the suitability of two types of deep autoencoders.
We report some interesting findings at the architectural level of the specific problem of modelling text at the sentence level.

The rest of the paper is structured as follows. 
A brief introduction to deep autoencoders is given in Section~\ref{sec: models}.
Section~\ref{sec: metrics} gives details about the analysis framework of the autoencoder learning, experiments and results.
The discussion on critical bottleneck dimensionality and an automatic way to estimate it is given in Section~\ref{sec: bottleneck dim}.
Finally, we present the conclusions and future directions of this work in Section~\ref{sec: conclusions}.

\section{Models}
\label{sec: models}
In this section we describe the models we have considered for performing dimensionality reduction of text data. First, we provide a brief introduction to autoencoders. Then, in sub-section 2.1, we present the binary deep autoencoder model (bDA); and, in sub-section 2.2, we describe the replicated softmax deep autoencoder (rsDA). Finally, in sub-section 2.3, we discuss the training procedure in detail. 

Both of the considered models differ in the way they model the text data. 
While the binary deep autoencoder models the presence of the term into the document (\textit{binary}), the replicated softmax deep autoencoder directly models the count of the term (i.e., \textit{term frequency}) in the document.

The autoencoder is indeed a network which tries to learn an approximation of the identity function so as the output is similar to input.
The input and output dimensions of the network are the same ($n$).
The autoencoder approximates the identity function in two steps: \textit{i)} reduction, and \textit{ii)} reconstruction.
The reduction step takes the input $x \in \mathbb{R}^n$ and maps it to $y \in \mathbb{R}^m$ where $m<n$ which can be seen as a function $y=g(x)$ with $ g: \mathbb{R}^n\to\mathbb{R}^m$.
On the other hand, the reconstruction step takes the output of the reduction step $y$ and maps it to $\hat{x} \in \mathbb{R}^n$ in such a way $\hat{x} \approx x$ which is considered as a $\hat{x}=f(y)$ with function $f:\mathbb{R}^m\to\mathbb{R}^n$.
The full autoencoder can be seen as $f(g(x))\approx x$.

\begin{figure}[!ht]
\centering
\includegraphics[scale=0.8]{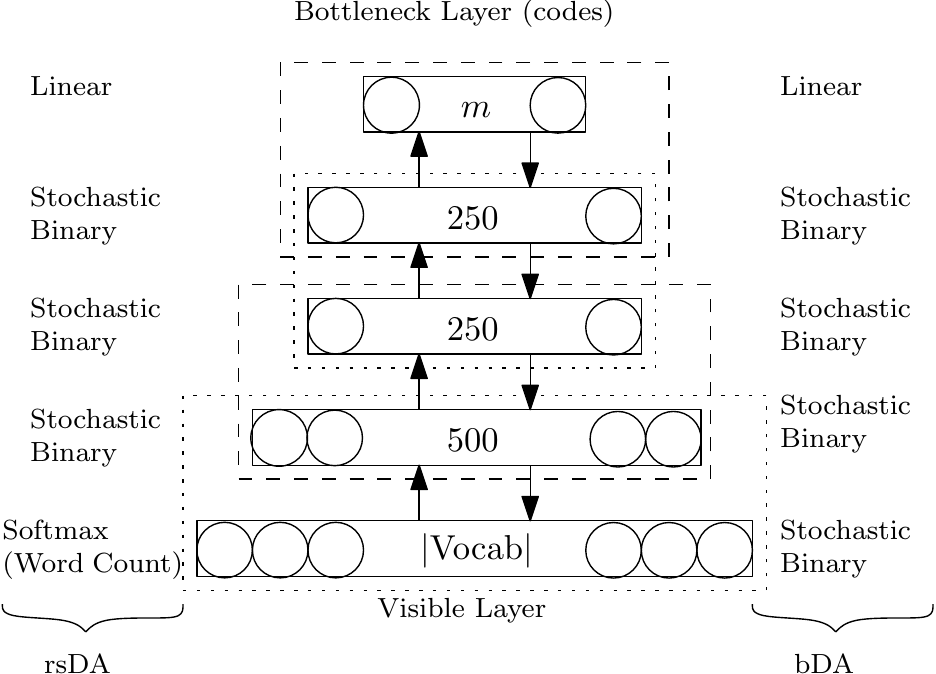}
\caption{Architecture of the deep autoencoders. The binary and replicated softmax deep autoencoders are denoted as bDA and rsDA. $|$Vocab$|$ is the size of vocabulary at the input layer.}
\label{fig: arch}
\end{figure}

In a neural network based implementation of the autoencoder, the visible layer corresponds to the input $x$ and the hidden layer corresponds to $y$.
There are two variants of autoencoders: \textit{i)} with a single hidden layer, and \textit{ii)} with multiple hidden layers.
If there is only one single hidden layer, the optimal solution remains the PCA projection even with the added non-linearities in the hidden layer~\cite{bourlard:1988}.
The PCA limitations are overcome by stacking multiple encoders, constituting what is called a deep architecture. This deep construction is what leads to a truly non-linear and powerful reduced space representation~\cite{hinton:2006}.
The deep architecture is constituted by stacking multiple restricted boltzmann machines (RBM) on top of each other as shown in Fig.~\ref{fig: arch}.

Each RBM is a two-layer bipartite network with a visible layer ($\textbf{v})$ and a hidden layer ($\textbf{h}$).
Both layers are connected through symmetric weights ($\textbf{w}$).
Usually the hidden units correspond to \textit{latent} variables.
It is very easy to sample the data from visible to hidden layer and vice-versa.
The two models we consider here, bDA and rsDA, primarily differ in the bottom-most RBM, i.e. the way they model the input data.
In a nutshell, in the case of bDA, the bottom-most RBM is a standard RBM with stochastic binary (visible and hidden) layers; while, in the case of rsDA, the bottom-most RBM is based on the replicated softmax model (RSM)~\cite{salakhutdinov:2009}.

\subsection{Stochastic Binary RBM}

Stochastic binary RBMs have both, visible and hidden, layers as stochastic binary with sigmoid non-linearity.
Let visible units $\textbf{v} \in \{0,1\}^n$ be binary bag-of-words representation of text documents and hidden units $\textbf{h} \in \{0,1\}^m$ be the hidden latent variables.
The energy of the state $\{\textbf{v},\textbf{h}\}$ is as follows,
{
\begin{equation}
 E(\textbf{v},\textbf{h}) = -\sum_{i=1}^n a_iv_i - \sum_{j=1}^m b_jh_j - \sum_{i,j} v_ih_jw_{ij} 
\end{equation}
}
\vspace{-3mm}

\noindent where $v_i, h_j$ are the binary states of visible unit $i$ and hidden unit $j$, $a_i, b_j$ are their biases and $w_{ij}$ is the weight between them. 

Then, it becomes easy to sample the data in both directions as shown below,
{
\begin{equation}
    \hspace{-5mm} p(v_i=1|\textbf{h}) = \sigma (a_i+ \sum_{j} h_j W_{ij})
\end{equation}
}
\vspace{-3mm}
{
\begin{equation}
    \hspace{-5mm} p(h_j=1|\textbf{v}) = \sigma (b_j+\sum_{i} v_i W_{ij})
\end{equation}
}
\noindent where $\sigma(x) = 1/(1+\text{exp}(-x))$ is the logistic sigmoid function. 

\subsection{Replicated Softmax RBM}

The Replicated Softmax RBM is based on the Replicated Softmax Model (RSM) proposed by Salakhutdinov and Hinton~\shortcite{salakhutdinov:2009}. 

Let $\textbf{v} \in \{1,\hdots,K\}^D$, where $K$ is the vocabulary size, $D$ is size of the document and let $\textbf{h}\in\{0,1\}^m$ be stochastic binary hidden latent variables.
Considering a document with length $D$ , the energy of the state $\{\textbf{v},\textbf{h}\}$ is defined as,
{\small
\begin{equation}
 E(\textbf{v},\textbf{h}) = -\sum_{k=1}^K\hat{v}^ka^k -D\sum_{j=1}^m b_jh_j -\sum_{k,j} W_j^kh_j\hat{v}^k
\end{equation}
}

\noindent where, $\hat{v}=\sum_{i=1}^D v_i^k$ denotes the count data for the $k^{th}$ term. 

In RSM, the visible layer is softmax with multinomial visible units which represents the probability distribution of the word-count.
It is sampled $D$ times by using multinomial sampling to recover the original word-count data.
Another distinction of this model is scaling of the bias terms of the hidden layer which gives a way to handle the documents of different lengths.
In this case, the visible and hidden units are updated as shown below,

{
\begin{equation}
  p(v_i^k=1|\textbf{h}) = \frac{\text{exp}(b_i^k + \sum_j h_jW_{ij}^k)}{\sum_{q=1}^K \text{exp}(b_i^q+\sum_jh_jW_{ij}^q)}  
\end{equation}
}
\vspace{-3mm}
{
\begin{equation}
     p(h_j=1|\textbf{V}) = \sigma (a_j+\sum_{i=1}^D \sum_{k=1}^K v_i^k W_{ij}^k)
\end{equation}
}

\subsection{Training of Autoencoders}
\label{sec: training}
Autoencoders are typically trained in two steps: \textit{i)} greedy layerwise pre-training, and \textit{ii)} fine-tuning of the parameters to learn the identity approximation of the input data.
\subsubsection{Pre-training}
In this step, each RBM is trained greedily using contrastive divergence (CD) learning~\cite{hinton:2002}.
Here the RBMs are trained one by one starting from the bottom-most RBM.
The bottom-most RBM directly takes the input data while the upper RBMs take the output $p(\textbf{h}|\textbf{v})$ of the RBM below which is already trained.
We use the structure of the autoencoder 500-250-250-m as shown in Fig.~\ref{fig: arch}.
We train each RBM using CD$_1$ learning for 50 epoch where CD$_1$ refers to CD with 1 step of alternating Gibbs sampling~\cite{hinton:2002}.
\subsubsection{Fine-tuning}
\begin{figure}[!ht]
 \centering
 \includegraphics[scale=1]{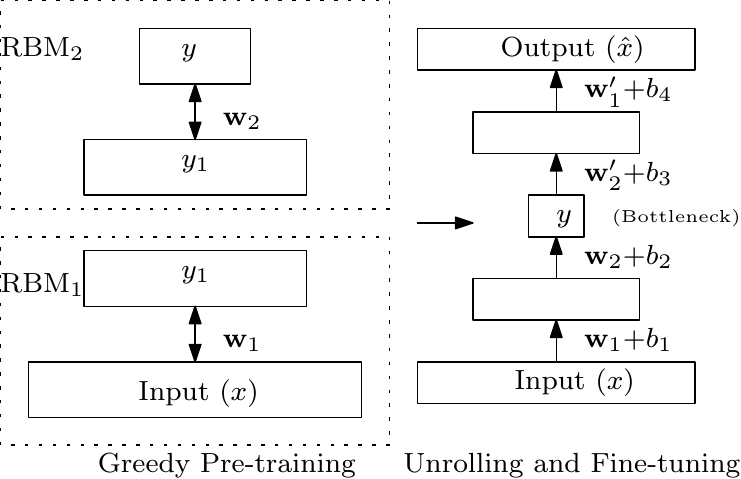}
\caption{\textbf{Left panel}: pre-training the stacked RBMs where upper RBMs take output of the lower RBM. \textbf{Right panel}: After pre-training the structure is ``unrolled'' to create a multi-layer autoencoder which is fine-tuned by backpropagation to perform $\hat{x}\approx x$.}
\label{fig: training}
\end{figure}

Once the RBMs are trained layer-wise, the autoencoder is unrolled as shown in Fig.~\ref{fig: training}.
The stochastic binary activities of the feature layers is replaced by the real-valued probabilities and the input data is backpropagated through the network to fine-tune the parameters of the entire network.
We calculate the cross-entropy error ($e$) between $\hat{x} = f(g(x))$ and $x$ as shown below and backpropagate it through the entire network.
\begin{equation}
 e = - \sum_{i=1}^n [x_i \log(\hat{x}_i) + (1-x_i) \log(1-\hat{x_i})]
\end{equation}
In case of bDA the binary input data is used to calculate $e$.
While for rsDA, the word-count input vectors are divided by the document length ($D$) to represent probability distribution which together with softmax output layer is used to calculate $e$.
\section{Qualitative Analysis and Metrics}
\label{sec: metrics}
In this section we describe the proposed metrics used for comparing the bDA and rsDA models. Subsequently, we present the comparative analysis of the two models.

The quality of the projections and the sufficiency of dimension $m$ are measured by the autoencoder's reconstruction ability.
Unfortunately, the mean squared error between the input $x$ and its reconstruction $\hat{x}$, referred as \textit{reconstruction error}, is a poor measure to estimate it.
It neither gives any details about the quality of the reconstructions in terms of text data representation nor the degree to which the structure of the data (distance between documents in original space) is preserved in the reconstruction space. 
Moreover, it is difficult to justify the adequacy of bottleneck dimension $m$ by simply using the \textit{reconstruction error}.

In literature, when autoencoders are used for dimensionality reduction for text data, most of the time, the quality of the algorithm is measured in terms of the accuracy of the end-task which may be text categorisation~\cite{hinton:2006}, information retrieval~\cite{salakhutdinov:2009b}, or topic modeling~\cite{salakhutdinov:2009}.
A shortcoming of this approach is that there is no way to estimate the full potential, or the upper bound, of the algorithm performance.
On the other hand, in the case of poor results, it becomes tougher to decide whether the training was proper or not.

As already mentioned before, in this work we propose two new metrics: \textit{i) structure preservation index (SPI)}, and \textit{ii) similarity accumulation index (SAI)}, which are intended to capture different aspects of the autoencoder's reconstruction capability, like the structural distortion and semantic similarity of the reconstructed vectors with respect to the original ones.
Considering these two metrics, along with the \textit{reconstruction error}, allows for a much better assessment of confidence regarding the quality of the network training process and its performance.

\paragraph{Structure Preservation Index (SPI):}
Consider the input data as $X$ where each row $x_i$ corresponds to the vector space representation of the $i^{th}$ document and $\hat{X}$ is its corresponding reconstruction.
$X$ and $\hat{X}$ are $p\text{x}n$ matrices where $p$ is the total number of documents in the input data and $n$ is the vocabulary size.
Compute matrix $D$ for $X$ such that $D_{ij}$ is the cosine similarity score between $i^{th}$ and $j^{th}$ rows of $X$.
Similarly calculate $\hat{D}$ for $\hat{X}$.
$D$ and $\hat{D}$ can be seen as similarity matrices of the original data and its reconstruction, respectively, where $D_{ij}=\hat{D}_{ij}=1,\forall i=j$.
The SPI is calculated as follows,
{
\begin{equation}
 \text{SPI} = \frac{1}{p^2} \sum_{ij}||D_{ij}-\hat{D}_{ij}||^2
\end{equation}
}
\noindent Notice that according to this definition, SPI captures the structural distortion incurred by the $f(g(X))$ process.
Ideally, SPI should be zero.

\paragraph{Similarity Accumulation Index (SAI):}
Different from SPI, which assesses structural distortion, SAI attempts to capture the quality of the reconstructed vectors by measuring the cosine similarity between each original vector and its reconstructed version. Indeed, this verifies the preservation of the relative strength of the vector-dimensions in the reconstruction.

SAI is computed by the normalised accumulation of cosine similarities between each input document and its reconstruction.
Ideally, SAI should be one. 
{
\begin{equation}
 \text{SAI} = \frac{1}{p} \sum_{i=1}^p \text{cosine}(x_i,\hat{x}_i)
\end{equation}
}
\subsection{Comparative Evaluation of Models}
\label{sec: data}
We carried out an experiment of dimensionality reduction for text sentences, where data sparseness plays a more critical role than in the case of full documents (dimensionality reduction applied to full documents is the case that has been mostly explored in the literature).

In this study we aim at applying autoencoder techniques at the level of sentences to open its way for sentence-centered applications, such as machine translation, text summarization and automatic dialogue response.

For this experiment, we used the Bible dataset, which contains 25122 training and 995 test sentences.
All sentences were processed by a term-pipeline of stopword-removal and stemming which is referred as $\textbf{Vocab}_1$.
After that we kept only those terms which were non-numeric, at least 3-characters long and appeared in at least 5 training sentences. We refer to this filtered vocabulary as $\textbf{Vocab}_2$.
For English partition of the dataset, $\textbf{Vocab}_1$ and $\textbf{Vocab}_2$  are 8279 and 3100 respectively.

Next, we present the results for English using both models, bDA and rsDA, and present the qualitative analysis of the reconstructions with the help of the aforementioned metrics.
We train both autoencoders with the structure 500-250-250-40 as described in Section~\ref{sec: training}.
The results are presented in Table~\ref{tab: both autoencoders}.
\begin{table}[!ht]
\centering
 \begin{tabular}{|l|ccc|}
  \hline
  \textbf{Model} & \texttt{\textbf{RC}} & \texttt{\textbf{SPI}} & \texttt{\textbf{SAI}}\\ \hline
  rsDA (\textit{pt}) & 0.1192 & 0.7258 & 0.2132 \\
  rsDA (\textit{bp}) & 0.0834 & 0.0049 & 0.5768 \\ \hline
  bDA (\textit{pt}) & 8.0012 & 0.0712 & 0.3528 \\
  bDA (\textit{bp}) & 5.4829 & 0.0035 & 0.6667 \\ \hline
 \end{tabular}
 \caption{The performance of bDA and rsDA in terms of different metrics. \texttt{\textbf{RC}} denotes \textit{reconstruction error} while \textit{pt} and \textit{bp} denote if the model is only pre-trained and fine-tuned after pre-training, respectively.}
 \label{tab: both autoencoders}
\end{table}

\subsection{Analysis and Discussion}
When operating in vector space, it is important to understand the amount of distortion incurred by the network on the structure of the data during the process of $f(g(x))$.
The network uses the \textit{reconstruction error} during the training to update parameters but it does not give much insight about the quality of the network.
One more limitation of the \textit{reconstruction error} is that it is not bounded and not comparable across different models $e.g.$ bDA and rsDA.
The \textit{reconstruction error} is calculated between the softmax output and the probability distribution of terms in case of rsDA hence it is not comparable to that of bDA (see Table~\ref{tab: both autoencoders}).

The two proposed metrics, SPI and SAI are both bounded by [0,1] and comparable across the models.
SPI gives the measure of how the similarity structure of sentences among each other is preserved in the reconstruction space which in turn gives a measure of trustworthiness of the network for similarity estimation.
Although both models show descent performance in terms of SPI after backpropagation, bDA is 28.57\% better than rsDA in terms of SPI.

\begin{figure}
 \begin{tabular}{c}
  \includegraphics[scale=0.5]{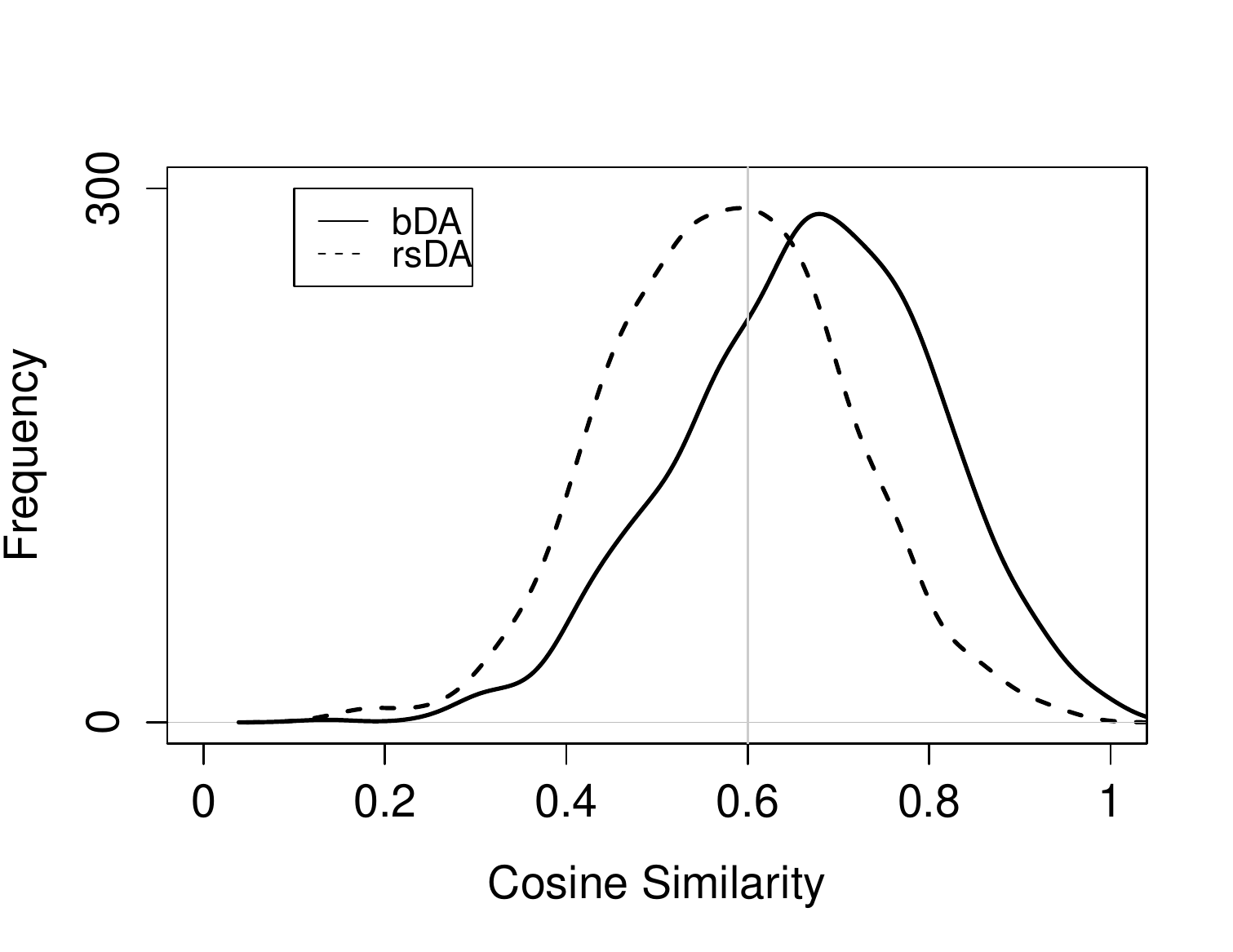} \\
 \end{tabular}
\caption{Histogram of cosine similarity between test samples and their reconstructions for bDA and rsDA.}
\label{fig: histogram}
\end{figure}

It is also important to assess the similarity between each input vector and its corresponding reconstruction which is captured by SAI.
In terms of SAI, bDA is 15.59\% better than rsDA.
This is better understood in Fig.~\ref{fig: histogram}, where it can be noticed that, in the case of rsDA, for more than half of the test samples cosine similarity with their reconstruction is $\leq 0.6$.
Although rsDA has been reported in the literature to better perform at the document level, our results demonstrate that bDA is a more suitable model to be used when using autoencoder representations at the sentence level. This can be explained by the fact that rsDA uses multinomial sampling to model the word-count, which happens not to be suitable at the sentence level for two reasons: \textit{i)} most of the terms appear only once in the sentences, and \textit{ii)} sampling the distribution of terms $D$ times is less reliable when $D$ is quite small which is the case in sentences compared to full documents.

Finally, as argued by Erhan et al.~\shortcite{erhan:2010}, pre-training helps to initialise the network parameters in a region to find optimal solution.
It can clearly be noticed that pre-training is necessary but itself is not enough to put aside backpropagation.

\section{Critical Bottleneck Dimensionality}
\label{sec: bottleneck dim}
In this section we present the analysis on the adequacy of the size of bottleneck layer.
\begin{figure*}[!ht]
 \centering
\begin{tabular}{ccc}
 \includegraphics[scale=0.35]{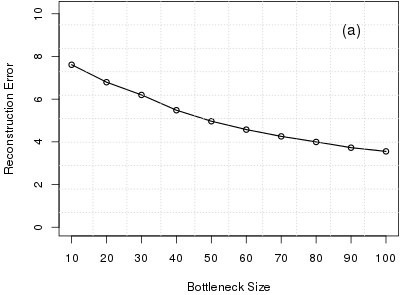} &
 \includegraphics[scale=0.35]{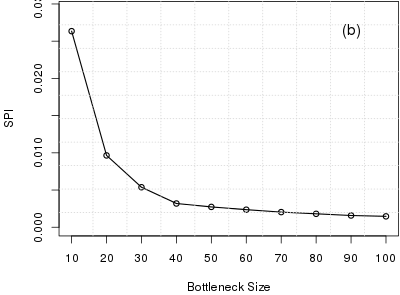} &
 \includegraphics[scale=0.35]{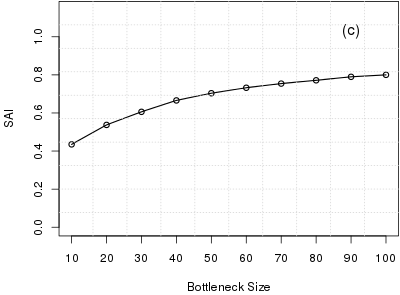} \\
\end{tabular}
\caption{\textit{Reconstruction error} \textit{(a)}, SPI \textit{(b)} and SAI \textit{(c)} metrics while squeezing the bottleneck layer from 100 to 10 of bDA.}
\label{fig: bottleneck-recon-spi-comparison}
\end{figure*}
The top-most hidden layer of an autoencoder is commonly referred to as the bottleneck layer.
The reconstruction ability of the autoencoder is highly related to the size of the bottleneck layer, in the sense that the smaller the size of the bottleneck layer is, the higher the loss of information is.

The reduction step of autoencoders is also called \textit{hashing}, and because similar sentences in the projected space are near to each other, this technique is also referred to as \textit{semantic hashing}.
It is important to choose a proper size of the bottleneck layer because of two reasons: \textit{i)} a too large size may lead to redundant dimensions and high computational cost, and \textit{ii)} a too small size might lead to high information loss.

The adequacy of the bottleneck dimension, which we refer to as critical bottleneck dimensionality here, is rarely addressed in the literature.
In this section of the study, we present an analysis on the effects of choosing different sizes for the bottleneck layer, as well as we provide an empirical method to choose the critical bottleneck dimensionality properly.

\subsection{Metric Selection}
\label{sec: bottleneck experiment}
We squeeze the bottleneck layer of the autoencoder to identify whether there was a dimensionality region at which the \textit{reconstruction error}, SPI and SAI metrics exhibit a clear change in its behaviour.
Typically, this region is referred to as the ``elbow region''.
We trained the autoencoder varying down the size of the bottleneck layer from 100 to 10 with step-sizes of 10.
Fig.~\ref{fig: bottleneck-recon-spi-comparison} shows the values of \textit{reconstruction error}, SPI and SAI for different sizes of bottleneck layer. 

As it becomes evident from the figure~\ref{fig: bottleneck-recon-spi-comparison}, SPI is the metric exhibiting the clearest ``elbow region'' pattern, hence we will use this metric for determining the critical bottleneck dimensionality. Indeed, it can be noticed that both the \textit{reconstruction error} and SAI show a quasi-linear behaviour with almost constant slope, while SPI clearly captures that below $m=40$, the network starts losing the structural information within the data.
This result shows that care must be taken to select a proper bottleneck dimension and it is important not to choose the bottleneck dimension below the point where SPI changes its behaviour.

\subsection{Identification of Critical Dimensionality}
To identify the critical bottleneck dimensionality, we calculated the percentage difference between the slopes connecting consecutive bottleneck sizes in the SPI curve.
This captures the point in the ``elbow region'' at which the slope of the SPI curve is steepest.
Consider three points in SPI plot: $a_1$, $a_2$ and $a_3$.
Let $s_1^2$ and $s_2^3$ be the slopes of lines connecting $a_1-a_2$ and $a_2-a_3$, respectively.
Then the percentage difference between $s_1^2$ and $s_2^3$ gives the steepness of the curve at point $a_2$.
We calculate this figure for every point in the range and estimate the \textit{critical dimensionality} at which the percentage difference is the largest.
This method enables us to automatically find the adequate bottleneck dimension. The algorithmic implementation of this method is described in Fig.~\ref{fig: algo}.
\begin{figure}[!h]
\centering
 \begin{tabular}{|p{7cm}|}
  \hline
  \textbf{Method}: Estimation of \textit{critical} dimension\\
  \hline
  \textbf{Input}: $ A, B$\\
  \textbf{Output}: $C$\\
  $A$ = set of bottleneck dimensions \\
  $B$ = set of SPI values, where $b_i$ = SPI$(a_i) \in A$ \\
  $C$ = set of steepness values at each point\\
  \textbf{for each} $a_{i-1}, a_i, a_{i+1} \in A$ \\
  \hspace{3mm} \textbf{get} $b_{i-1}, b_i, b_{i+1} \in B$\\
  \hspace{3mm} \textbf{calc.} $s_{i-1}^i$, $s_i^{i+1}$ where,\\
  \hspace{8mm} $s_{i-1}^i$ = slope(($a_{i-1},b_{i-1}$), ($a_i,b_i$))\\
  \hspace{3mm} \textbf{calc.} $c_i$ = \% diff ($s_{i-1}^i$,$s_i^{i+1}$) \\
  \hspace{3mm} \textbf{add} $c_i$ to $C$\\
  \textbf{end}\\
  \textbf{plot} $C$\\
  \textit{critical dim.} = right-most large peak\\
  \hline
 \end{tabular}

\caption{Method to identify the \textit{critical dimensionality} for the bottleneck layer.}
\label{fig: algo}
\end{figure}

\section{Conclusions and Future Research Directions}
\label{sec: conclusions}
In this work we have presented a comprehensive study on the use of autoencoders for modelling text data, in which differently from previous studies we focused our attention in the following issues:
\begin{itemize}
\item{We explored the suitability of two different models bDA and rsDA for constructing deep autoencoder representations of text data at the sentence level.}
\item{We proposed and evaluated two novel metrics which assess the reconstruction quality of an autoencoder with regards to the particular problem of text data representation.}
\item{We proposed an automatic method to find the critical bottleneck dimensionality for text language representation, below which structural information is lost.}
\end{itemize}

As a result of this study we have found that the bDA model is most suitable for constructing and training autencoders for handling text data at the sentence level. 
We also found that our defined SPI (Structure Preservation Index) metric allows for a better discrimination and identification of the critical bottleneck dimensionality.

As future work, we want to study the suitability of our proposed metrics, especially SPI, as error metric during the autoencoder fine tuning stage. If this metric can be used along with back-propagation, we envisage a new generation of text-oriented autoencoders able to provide a much better characterization of the linguistic phenomenon in text data. 

\section*{Acknowledgment}
The work of the first and third authors was carried out in the framework of the WIQ-EI IRSES
project (Grant No. 269180) within the FP 7 Marie Curie, the DIANA APPLICATIONS Finding
Hidden Knowledge in Texts: Applications (TIN2012-38603-C02-01) project and the VLC/CAMPUS
Microcluster on Multimodal Interaction in Intelligent Systems.

\bibliographystyle{acl}
\bibliography{reference}
\end{document}